\documentclass[onecolumn,prb,aps,amsfonts,amssymb,amsmath]{revtex4}
\usepackage{graphicx}
\usepackage{hyperref}
\usepackage{color}
\begin{document}

%------------------------------------------------------
\title{Wigner and Kondo physics in quantum point contacts \\ revealed by scanning gate microscopy}
%------------------------------------------------------

\author{B. Brun$^{1,2}$, F. Martins$^3$, S. Faniel$^3$, B. Hackens$^3$, G. Bachelier$^{1,2}$,
A. Cavanna$^4$, C. Ulysse$^4$, A. Ouerghi$^4$, U. Gennser$^4$, D. Mailly$^4$, S. Huant$^{1,2}$, V.
Bayot$^{1,3}$, M. Sanquer$^{1,5}$ \& H. Sellier$^{1,2\ast}$}

\affiliation{
$^1$Univ. Grenoble Alpes, F-38000 Grenoble, France \\
$^2$CNRS, Inst. NEEL, F-38042 Grenoble, France \\
$^3$IMCN/NAPS, Universit\'e catholique de Louvain, B-1348 Louvain-la-Neuve, Belgium \\
$^4$CNRS, Laboratoire de Photonique et de Nanostructures, UPR20, F-91460 Marcoussis, France \\
$^5$CEA, INAC-SPSMS, F-38054 Grenoble, France \\
$^\ast$Correspondence should be addressed to H.S. (hermann.sellier@neel.cnrs.fr). }

%\date{\today}

\begin{abstract}
Quantum point contacts exhibit mysterious conductance anomalies in addition to well known
conductance plateaus at multiples of $2e^2/h$. These 0.7 and zero-bias anomalies have been
intensively studied, but their microscopic origin in terms of many-body effects is still highly
debated. Here we use the charged tip of a scanning gate microscope to tune \textit{in situ} the
electrostatic potential of the point contact. While sweeping the tip distance, we observe
repetitive splittings of the zero-bias anomaly, correlated with simultaneous appearances of the 0.7
anomaly. We interpret this behaviour in terms of alternating equilibrium and non-equilibrium Kondo
screenings of different spin states localized in the channel. These alternating Kondo effects point
towards the presence of a Wigner crystal containing several charges with different parities.
Indeed, simulations show that the electron density in the channel is low enough to reach
one-dimensional Wigner crystallization over a size controlled by the tip position.
\end{abstract}

\maketitle

%------------------------------------------------------
\section{Introduction}
%------------------------------------------------------

Quantum point contacts~\cite{van-Wees-1988} (QPCs) are among the simplest quantum devices made out
of a two-dimensional electron gas (2DEG). Applying a negative voltage on a split-gate creates a
quasi-one-dimensional (1D) channel connected to large 2D reservoirs. This narrow channel behaves as
an electron wave-guide and transmits a finite number of modes, each of them carrying one quantum of
conductance $G_0=2e^2/h$ ($e$ is the electron charge and $h$ the Planck constant). As a result, the
conductance versus gate voltage curve shows a series of quantized plateaus with transitions which
are well reproduced by a single-particle model~\cite{Buttiker-1990}.

However, since the early days of QPCs, a shoulder-like feature is commonly
observed~\cite{Thomas-1996} at a conductance around 0.7~$G_0$, which cannot be explained by
single-particle theories. With lowering temperature, this ``0.7 anomaly'' rises to reach the first
plateau, and a zero-bias peak called ``zero-bias anomaly'' (ZBA) emerges in the non-linear
differential conductance~\cite{Cronenwett-2002}. These anomalies have been extensively studied
through transport
experiments~\cite{Thomas-1996,Kristensen-2000,Cronenwett-2002,Reilly-2002,Hew-2008}, revealing the
complexity of the underlying phenomena. Different theoretical models have been
proposed~\cite{Wang-1998,Spivak-2000,Sushkov-2001,Meir-2002,Matveev-2004,Rejec-2006}, but no
consensus could be reached so far on their interpretation~\cite{Micolich-2011}.

Recently, an experiment using several gates to vary the channel length~\cite{Iqbal-2013a} revealed
the possible existence of several emergent localized states responsible for the conductance
anomalies. At the same time, a different theoretical model was proposed~\cite{Bauer-2013},
explaining the anomalies without invoking localized states in the channel. As stressed in
Ref.~\cite{Micolich-2013}, investigating these anomalies using scanning probe techniques could make
it possible to check the existence of spontaneously localized states and discriminate between these
two proposals: this is the aim of the present letter.

Here we perform scanning gate microscopy~\cite{Eriksson-1996} (SGM), in which a negatively charged
tip is scanned above the sample surface and modifies the electrostatic potential in the 2DEG. This
local potential change induces electron back-scattering towards the QPC, which can be used to image
single-particle phenomena such as wave-function quantization in the channel~\cite{Topinka-2000},
branched flow in the disorder potential~\cite{Topinka-2001}, interference patterns induced by the
tip~\cite{Leroy-2005,Jura-2007,Kozikov-2013}, or to investigate electron-electron interactions
inside~\cite{Freyn-2008} or outside~\cite{Jura-2010} the QPC. This movable gate can also be used to
tune \textit{in situ} the saddle potential of the QPC, in a more flexible and less invasive way
than fixed surface gates, and probe intrinsic properties of the QPC such as the 0.7
anomaly~\cite{Crook-2006,Iagallo-2013}.

Here we show that approaching the tip towards the QPC produces an oscillatory splitting of the ZBA,
correlated with simultaneous appearances of the 0.7 anomaly, thereby confirming that both features
share a common origin~\cite{Cronenwett-2002,Iqbal-2013a}. We interpret these observations as the
signature of a small one-dimensional Wigner crystal~\cite{Wigner-1934,Schulz-1993,Deshpande-2008}
forming in the channel~\cite{Hew-2009} (a quantum chain of charges localized by Coulomb
interactions in absence of disorder). The number of charges in this many-body correlated state is
tuned by changing the tip position, leading alternatively to a single- or a two-impurity Kondo
effect (screening of a localized spin by conducting electrons), with a conductance peak either at
zero, or at finite bias, depending on the charge parity.

Our observations therefore strongly support the existence of emergent localized states, as
suggested in Ref.~\cite{Iqbal-2013a} where the number of localized charges is controlled by
changing the effective channel length. Here we show that a similar effect is observed when changing
the distance of an additional gate placed around the QPC. To understand this new result, we perform
classical electrostatic simulations and evaluate the size of the region where electrons should form
a 1D Wigner crystal thanks to the critically low electron density. We show that the calculated size
of this small crystal is in good agreement with the observed change in the number of localized
charges, thereby revealing that Wigner crystallization is, to our opinion, the correct way to
understand this spontaneous localization.

\begin{figure}[b]
\begin{center}
\includegraphics[width=16cm]{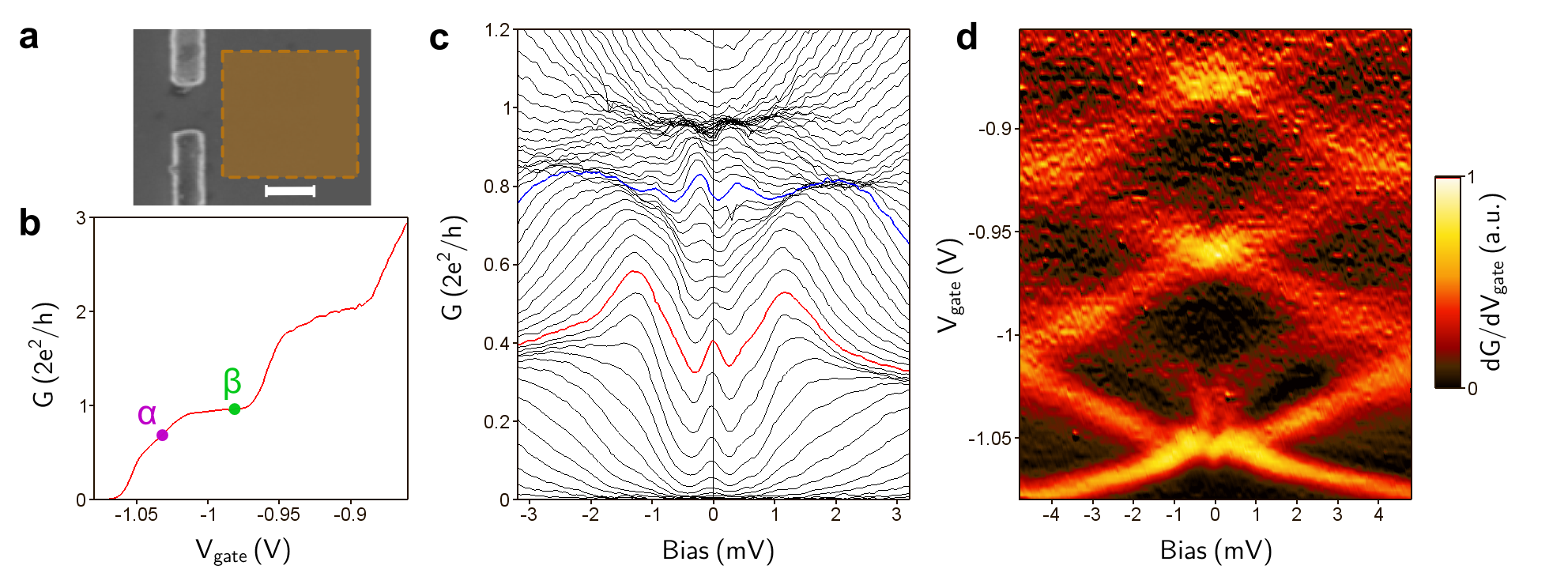}
\caption{\textbf{Transport measurements.} Base temperature is 20~mK. (a) Electron micrograph of the
QPC gates. The scale bar is 300~nm. The dashed box indicates the position of the scanning area used
in Fig.~2. (b) Differential conductance $G$ at zero bias versus split-gate voltage $V_{\rm gate}$.
The 0.7 anomaly is visible below the first plateau. Positions $\alpha$ and $\beta$ are used in
Fig.~2a,b. (c) Differential conductance $G$ versus source-drain bias for different gate voltage
$V_{\rm gate}$ from -1.08~V to -0.96~V. The zero-bias peak in the red curve splits into finite-bias
peaks in the blue curve. (d) Numerical derivative of the differential conductance ${\rm d}G/{\rm
d}V_{\rm gate}$ versus bias and gate voltage. Yellow lines highlight transitions between
conductance plateaus.}
\end{center}
\end{figure}

\begin{figure}[b]
\begin{center}
\includegraphics[width=14cm]{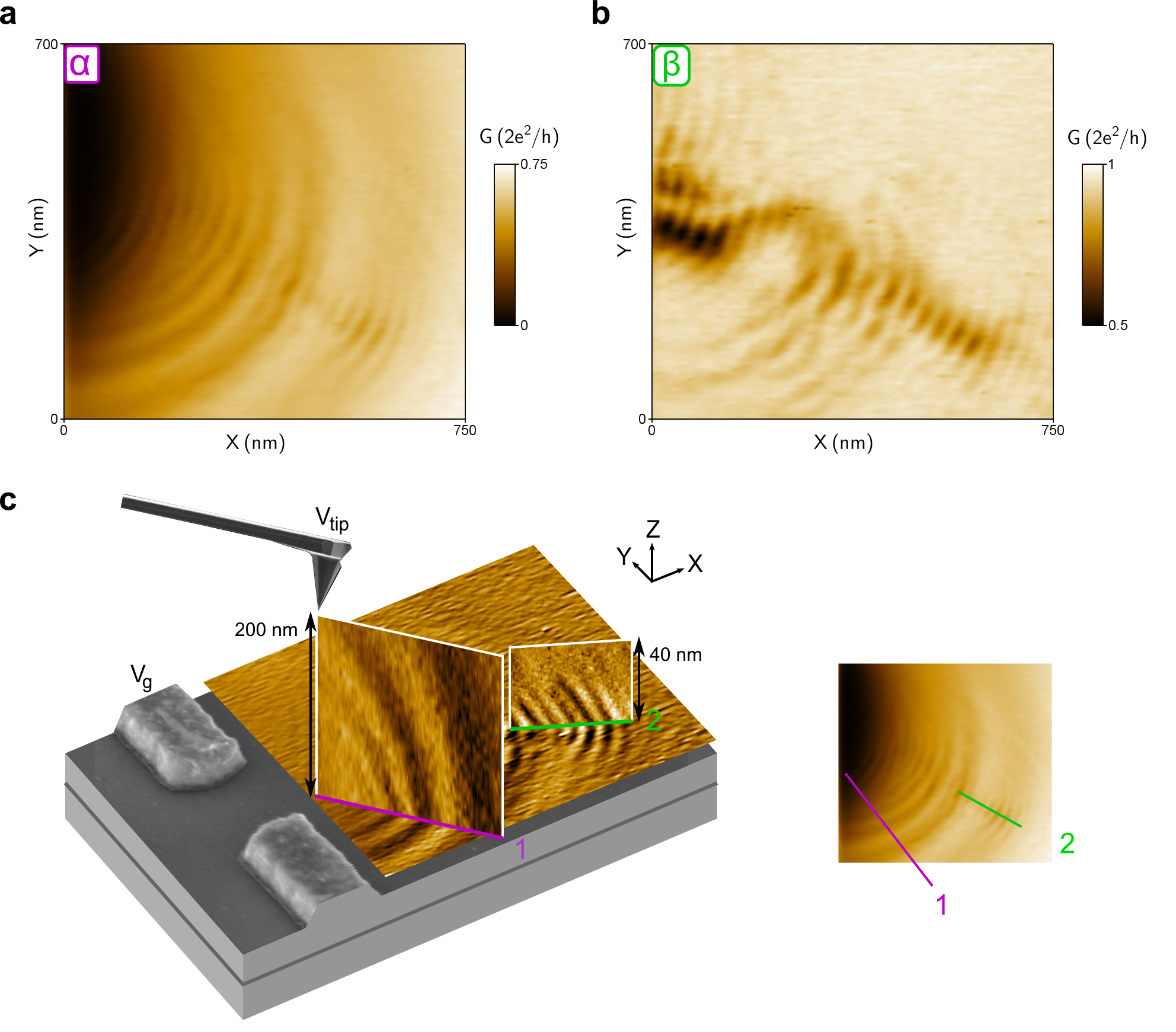}
\caption{\textbf{Scanning gate microscopy.} Base temperature is 20~mK. (a,b) SGM maps of the QPC
conductance $G$ versus tip position in the $(X,Y)$ horizontal plane for gate voltages $V_{\rm
gate}=-1$~V (a) and $-0.95$~V (b) corresponding respectively to points $\alpha$ and $\beta$ as
defined in Fig.~1b (gate voltages are shifted by 35~mV in presence of the tip). Concentric rings
are only visible at $\alpha$, and interference fringes are more contrasted at $\beta$. Additional
data are presented in Supplementary Figure 2. (c) Schematic view of the SGM experiment showing the
tip scanning above the 2DEG near the QPC gates and three SGM maps. The horizontal map is the same
as in (a), but the data have been differentiated with respect to the Y-coordinate to highlight
details. The two vertical maps are recorded in planes perpendicular to the surface along the purple
line 1 (size $500 \times 200$~nm, gate voltage $\alpha$) and the green line 2 (size $250 \times
40$~nm, gate voltage $\beta$) as indicated on the right image (identical to (a)). The two vertical
maps have been differentiated with respect to their horizontal coordinate to highlight details (raw
data are shown in Supplementary Figure 3). The vertical map along line 1 reveals that the
concentric rings visible in (a) form also rings in the vertical plane, whereas the vertical map
along line 2 shows that interference fringes disappear rapidly with the tip-to-surface distance.}
\end{center}
\end{figure}

\begin{figure}[b]
\begin{center}
\includegraphics[width=16cm]{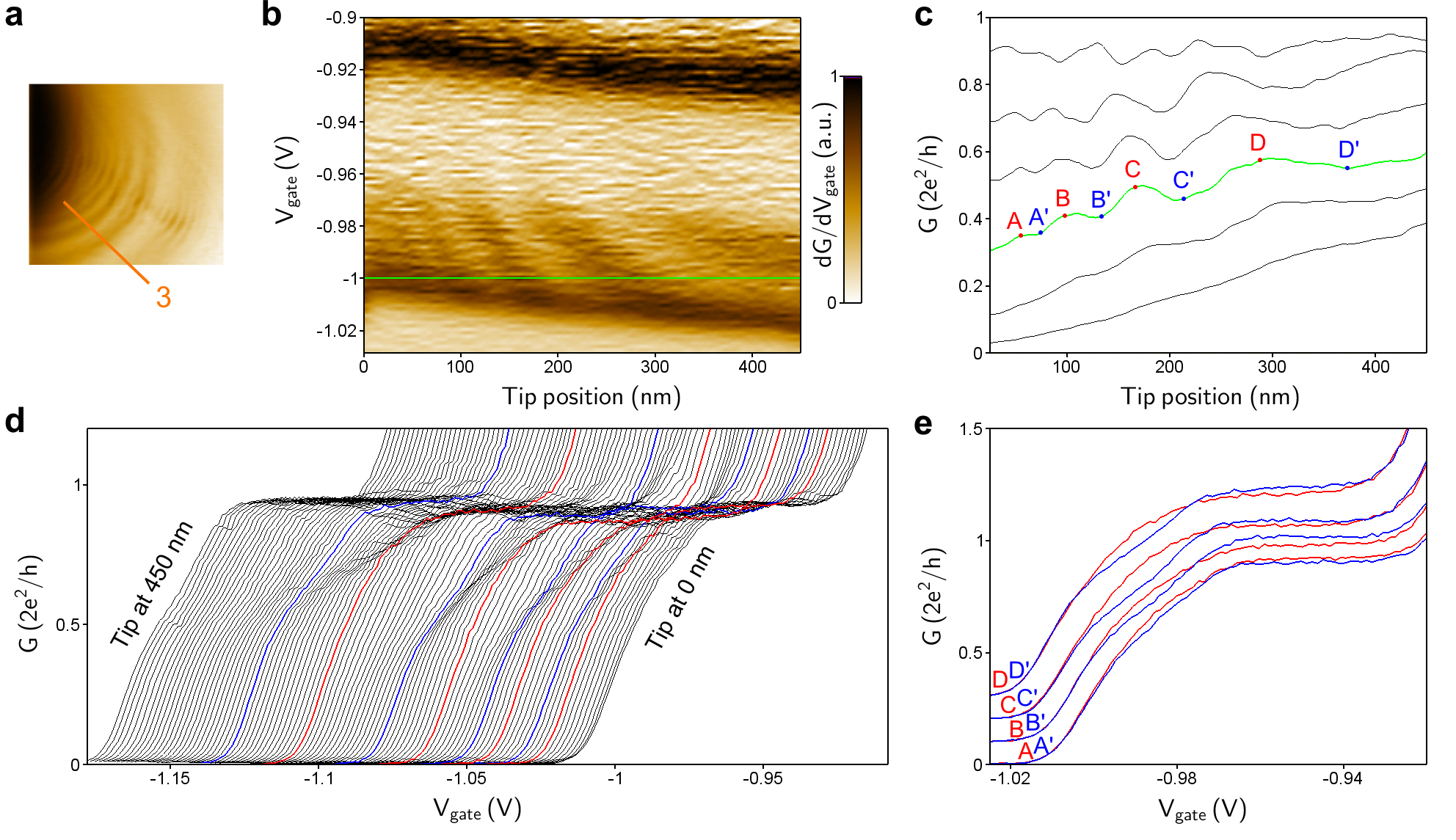}
\caption{\textbf{Modulation of the 0.7 anomaly.} The figure analyses the zero-bias conductance
oscillations when the tip is scanned along the orange line 3 indicated in (a) with the origin of
positions in the QPC direction. (b) Trans-conductance ${\rm d}G/{\rm d}V_{\rm gate}$ versus tip
position and gate voltage. Black regions correspond to transitions between plateaus. The
conductance oscillations are only visible below the first plateau. (c) Conductance $G$ versus tip
position for gate voltages $V_{\rm gate}=-0.964$~V, $-0.983$~V, $-0.992$~V, $-1.000$~V, $-1.006$~V,
$-1.012$~V (from top to bottom). Conductance extrema at $V_{\rm gate}=-1$~V (green curve) are
labelled A to D (maxima) and A' to D' (minima). The global slope corresponds to the rise of the
saddle-point potential when the tip approaches the QPC. (d) Conductance $G$ versus gate voltage for
different tip positions from 0 to 450~nm (successive curves are shifted to the left). (e) Same data
as in (d) but for tip positions A to D (red curves, shifted vertically) and A' to D' (blue curves,
shifted also horizontally to be compared with red curves). Red curves show no shoulder, whereas
blue curves show the 0.7 anomaly. Small differences between plateau values come from residual
interference fringes.}
\end{center}
\end{figure}

\begin{figure}[b]
\begin{center}
\includegraphics[width=16cm]{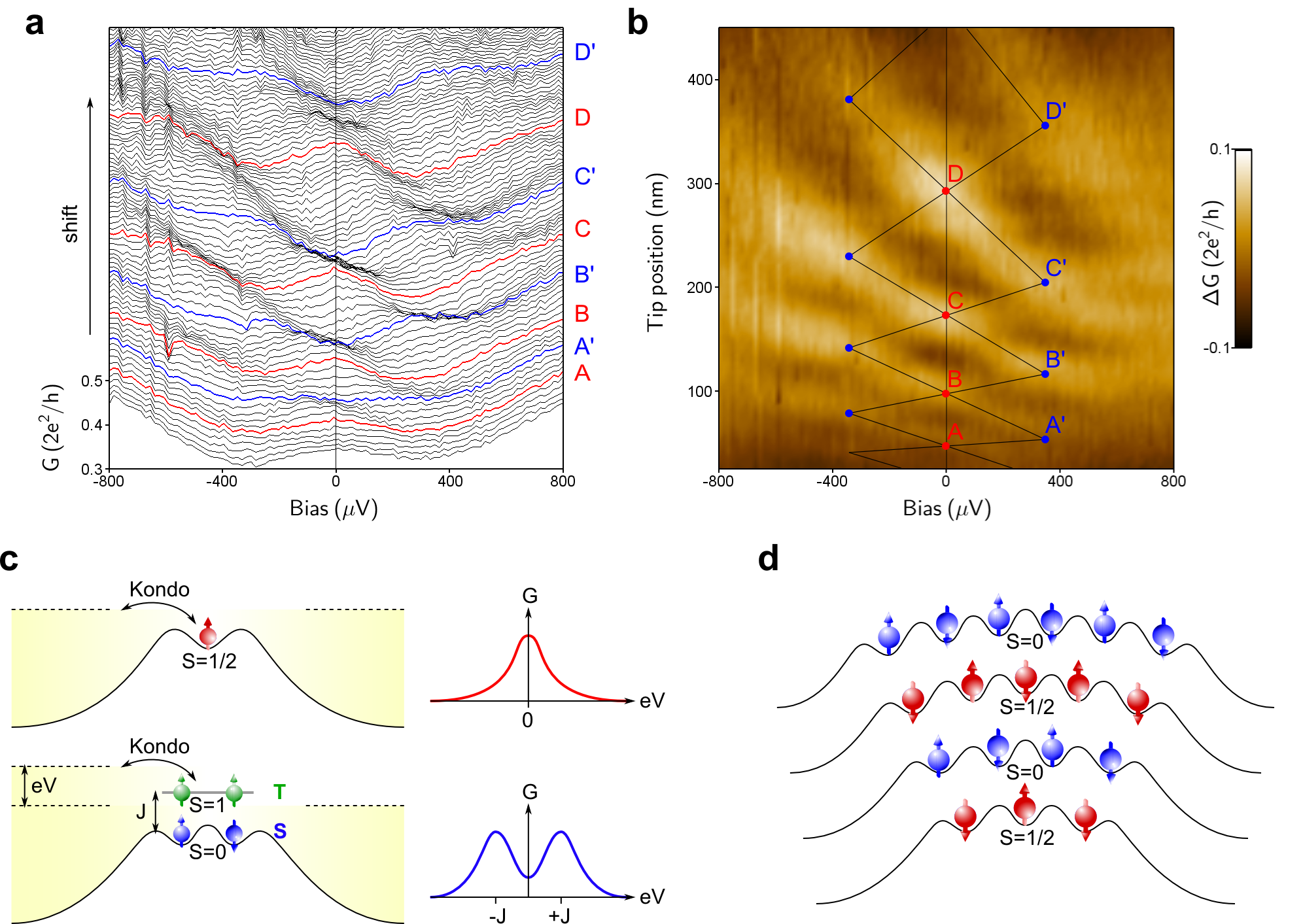}
\caption{\textbf{Successive splittings of the ZBA.} The figure analyses the low bias source-drain
spectroscopy when the tip is scanned along the orange line 3 indicated in Fig.~3a. (a) Differential
conductance $G$ versus source-drain bias at a fixed gate voltage $V_{\rm gate}=-1$~V for different
tip positions from 0 to 450~nm. Successive curves are shifted upwards by $0.0075\times 2e^2/h$.
Conductance peaks are visible at zero and finite bias on red and blue curves, respectively. (b)
Color plot of the same data as in (a) after subtraction of a smooth background to suppress the main
gating effect of the tip. Peak positions are indicated by dots. The successive ZBA splittings give
a checkerboard pattern. The asymmetry results from the bias-induced change of the QPC position. (c)
Schematic of the QPC potential with one (top) and two (bottom) localized electrons, corresponding
respectively to a $S=1/2$ ground state with a zero-bias Kondo peak and to a $S=0$ ground state with
finite-bias Kondo peaks involving the excited state $S=1$ with singlet-triplet energy splitting
$J$. The expected conductance $G$ versus bias $V$ is shown on the right for each state. (d)
Schematic of the QPC potential with an increasing number of electrons localized by Coulomb
interactions. The antiferromagnetic spin coupling in this small 1D Wigner crystal gives either a
$S=1/2$ ground state (ZBA) or a $S=0$ ground state (splitting of the ZBA), depending on the parity
(respectively odd or even).}
\end{center}
\end{figure}

%------------------------------------------------------
\section{Results}
%------------------------------------------------------

\subsection*{Transport measurements}

The QPC (see Methods and Fig.~1a) is cooled down to a temperature of 20~mK in a cryogenic scanning
probe microscope~\cite{Hackens-2010}. In the absence of the tip (moved several microns away), the
linear conductance shows the usual staircase behaviour versus gate voltage (Fig.~1b). The shoulder
below the first quantized plateau is the puzzling 0.7 anomaly. The source-drain bias spectroscopy
(Fig.~1c) shows that this shoulder evolves to a clear plateau at 0.85~$G_0$ at finite
bias~\cite{Patel-1991,Thomas-1998,Kristensen-2000}. The narrow peak around zero bias is the ZBA and
disappears above 1~K (Supplementary Figure 1). Its width of 200~$\mu$eV is much smaller than the 1D
subband spacing of 4.5~meV (Fig.~1d). Above 0.7~$G_0$, the ZBA splits into finite-bias
peaks~\cite{Iqbal-2013a,Komijani-2013} centred at $\pm$250~$\mu$V. We show in the following that
the presence of the 0.7 anomaly is related to this splitting of the ZBA.

\subsection*{Scanning gate microscopy}

When the tip is scanned near the QPC and polarized such as to deplete locally the 2DEG (see
Methods), we observe two distinct phenomena. On the first conductance plateau (Fig.~2b), SGM images
reveal the electron flow coming out of the QPC, with fringes spaced by half the Fermi wavelength,
as already observed by several groups~\cite{Topinka-2001,Jura-2007,Kozikov-2013}. The fringes
result from interferences of electrons backscattered by the depleted region below the tip and
reflected by impurities~\cite{Topinka-2001,Jura-2007} in the 2DEG or directly by the
gates~\cite{Leroy-2005,Jura-2009}.

Below the first plateau (Fig.~2a), SGM maps reveal a novel set of concentric rings centred on the
QPC, with a spacing increasing with tip distance (see also Supplementary Figure 2). As opposed to
the previous one-particle interference fringes, these new rings are not linked to the electron flow
(black region in Fig.~2b) but extend rather isotropically around the QPC, not only in the
horizontal plane but in all three directions of space. This is revealed by scanning the tip in a
vertical plane (Fig.~2c), unveiling half spheres centred on the QPC (purple line 1). This behaviour
contrasts with that of interference fringes (green line 2) that quickly disappear when the tip is
scanned more than 50~nm above the surface (see also Supplementary Figure 3). Interferences indeed
require electrons at the Fermi level to be backscattered by a depleted region below the tip, a
situation which is only obtained for the tip close enough to the 2DEG (and at low enough
temperature to avoid thermal averaging of the interferences). We therefore conclude that the new
rings are not interferences but result from a direct tuning of the electrostatic potential in the
QPC. The larger ring spacing at larger distances results from the smaller potential changes induced
by the tip.

\subsection*{Conductance anomalies}

To demonstrate that these rings correspond to modulations of the conductance anomalies, the tip is
scanned along a single line in a region with almost no interference (line 3 in Fig.~3a) and the QPC
parameters (gate and bias voltages) are varied. Fig.~3b shows that the ring-related conductance
oscillations are only visible for gate voltages in the transition below the first plateau, just
where the ZBA and 0.7 anomaly are observed. Fig.~3c shows how the conductance oscillations evolve
when the average conductance goes from 0 to $G_0$ while changing the gate voltage. The oscillations
are clearly visible between 0.4 and 0.8~$G_0$. They are blurred when approaching $G_0$ because some
interference fringes come into play. The increasing distance between conductance extrema (labeled A
to D for maxima and A' to D' for minima) is consistent with an oscillatory phenomenon in the QPC,
controlled by the decreasing electrostatic coupling to the tip. Plotting the conductance versus
gate voltage (Fig.~3d) reveals the oscillatory behaviour of the 0.7 anomaly. The amplitude of this
modulation can be read from Fig.~3e, where curves at positions X and X' are compared two-by-two
(curves are shifted horizontally to compensate for the drift of the pinch-off voltage while
approaching the tip). Curves at positions A to D are smooth with no shoulder, i.e. no anomaly,
whereas curves at positions A' to D' present a reduced conductance above 0.5~$G_0$, i.e. the 0.7
anomaly. The concentric rings observed in SGM images (Fig.~2a) therefore correspond to an
alternating modulation of the 0.7 anomaly when the tip approaches the QPC.

We now analyse the behaviour of the ZBA when the 0.7 anomaly repeatedly appears and disappears, and
show that both anomalies are linked. Fig.~4a shows the differential conductance versus source-drain
bias for different tip positions (same scan line as in Fig.~3a). Curves at positions A to D have a
peak centred at zero bias (ZBA), whereas curves at positions A' to D' have a dip at zero bias and
local maxima at $\pm$250~$\mu$V bias (splitting of the ZBA), on top of the same V-shaped
background. Scanning the SGM tip therefore produces a repetitive splitting of the ZBA, that draws a
checkerboard pattern in a color-plot of the spectroscopy versus tip position (Fig.~4b). Note that
the spontaneous splitting of the ZBA observed without the tip (Fig.~1c) also shows peaks at
$\pm$250~$\mu$V and probably has the same origin.

Considering the regularity of the concentric rings in Fig.~2a, this oscillatory behaviour of the
0.7 and zero-bias anomalies would be observed for any scanning line in a large range of angles (see
Supplementary Figure 4 and Supplementary Note 1). As a consequence, rings with conductance maxima
correspond to a simple staircase in the linear conductance and a ZBA in the non-linear
spectroscopy, whereas rings with conductance minima correspond to a 0.7 anomaly and a splitting of
the ZBA. This result shows that the ZBA suppresses the 0.7 anomaly at low
temperature~\cite{Cronenwett-2002} only if the ZBA is not split into finite-bias peaks.

%------------------------------------------------------
\section{Discussion}
%------------------------------------------------------

First, we would like to stress again that these new conductance oscillations cannot be explained by
interference effects in the 2DEG. One argument already given above is that interferences require
backscattering with a tip close to the surface, whereas the new rings are observed up to large tip
heights (Fig.~2c). A second argument is that interference fringes would have an increasing spacing
for short tip distances because the density is reduced close to the QPC and the electron wavelength
is larger, but the opposite behaviour is observed.

We now discuss a possible single-particle effect inside the QPC that, at first sight, could give
similar conductance oscillations. In case of a non-adiabatic transmission, wave-functions are
scattered by the QPC potential barrier and transmission resonances appear when the barrier length
is equal to an integer number of half the longitudinal wavelength. If the effect of the tip is to
change the channel length, such resonances could give conductance oscillations versus tip distance.
However, this single-particle mechanism cannot explain the repetitive splittings of the ZBA which
are simultaneous with the observed conductance oscillations and we therefore need another
explanation.

The ZBA in QPCs has been shown to scale with temperature and magnetic field like the Kondo effect
in quantum dots~\cite{Cronenwett-2002}. This effect corresponds to the screening of a single
degenerate level by a continuum of states, and therefore indicates the presence of a localized spin
in the QPC channel~\cite{Meir-2002}. Splittings of the ZBA have been observed recently in
length-tunable QPCs~\cite{Iqbal-2013a} and interpreted as a two-impurity Kondo
effect~\cite{Georges-1999,Aguado-2000}, involving non-equilibrium Kondo
screening~\cite{Lopez-2002,Kiselev-2003}, as commonly observed in quantum dots with even numbers of
electrons~\cite{Sasaki-2000}, coupled quantum dots~\cite{Jeong-2001}, and molecular
junctions~\cite{Roch-2008}.

We now consider different scenarios to explain the presence of such localized states in our system.
In a recent work on QPCs made out of a two dimensional hole gas, a spontaneous splitting of the ZBA
as the QPC opens has been reported~\cite{Komijani-2013}. This effect was attributed to a charge
impurity forming a potential well close to the channel, containing one or two charges, leading to
different types of Kondo screening. In our case, the spontaneous splitting of the ZBA as the QPC
opens (Fig.~1c) could be explained by this effect. However, the fact that approaching the tip
towards the QPC results in 4 successive splittings of the ZBA indicates that this impurity should
contain at least 8 charges, which is unlikely for a single impurity. Nevertheless, one could
imagine that a shallow quantum dot has formed in the QPC due to potential fluctuations induced by
residual disorder~\cite{Nixon-1991} and giving Coulomb blockade oscillations as often observed in
long 1D wires~\cite{Staring-1992}. The major argument to exclude this scenario is that the
split-gate has a larger capacitive coupling to the channel than the tip has (i.e. a larger
lever-arm parameter), so the split-gate should induce more charging events than the tip, but we observe
the opposite: approaching the tip by 600~nm produces four successive splittings of the ZBA and
sweeping the gate voltage produces only one splitting. It can therefore not be Coulomb blockade in
a disorder-induced quantum dot.

The only remaining possibility to explain the presence of localized states in the channel is a
spontaneous electron localization which is not induced by potential barriers but instead by
electron-electron interactions. Indeed, a large number of theoretical and numerical investigations
show that interactions can localize a finite number of electrons in the
channel~\cite{Sushkov-2003,Matveev-2004,Rejec-2006,Guclu-2009}. On the first conductance plateau
and below, transport can be considered as 1D, and the electron density is so low that the Coulomb
repulsion overcomes the kinetic energy. When the 1D density $n_{\rm 1D}$ fulfills the criterion
$n_{\rm 1D} \times a_{\rm B} < 1$, where $a_{\rm B}$ is the effective Bohr radius (10~nm in GaAs),
electrons are expected to spontaneously order in a crystal, with an inter-particle distance
minimizing Coulomb repulsion~\cite{Shulenburger-2008}. This many-body state, known as a Wigner
crystal~\cite{Wigner-1934,Schulz-1993}, has been suggested to be responsible for the 0.7 anomaly in
QPCs~\cite{Matveev-2004}. When the electron density in the channel is decreased below the critical
value, the density modulations evolve continuously from the $\lambda_{\rm F}/2$ periodicity of
Friedel oscillations to the $\lambda_{\rm F}/4$ periodicity of the Wigner
crystal~\cite{Soffing-2009}. Quantum Monte Carlo simulations have also shown that electrons in the
crystallized region can be relatively decoupled from the high density reservoirs and present an
antiferromagnetic coupling $J$ between adjacent spins~\cite{Guclu-2009}. In contrast to the case of
quantum dots with real tunnel barriers, electron localization in a QPC is not straightforward, and
results from emergent barriers in the self-consistent potential. On the other hand, the Kondo
effect requires a relatively open system with a good coupling to the reservoirs, and this makes the
QPC a suitable platform to observe Kondo phenomena on an interaction-induced localized state, as
shown recently in length-tunable QPCs~\cite{Iqbal-2013a}.

This last scenario being the most realistic one in our case, we therefore interpret the four
observed oscillations as a signature of eight successive states of a small non-uniform 1D Wigner
crystal with an alternating odd and even number of localized charges. Situations with an odd number
of electrons in a spin $S=1/2$ ground state show a ZBA due to Kondo screening of non-zero spin
states. Situations with an even number of electrons in a spin singlet $S=0$ ground state show a
splitting of the ZBA due to non-equilibrium Kondo screening~\cite{Lopez-2002,Kiselev-2003} of the
spin triplet $S=1$ excited state with peaks at a finite bias $eV=J$ (Fig.~4c). The four
oscillations, suggestive of eight successive states, reveal that a large number of electrons can
spontaneously localize in the channel of a QPC, as shown in Fig.~4d. Observing Kondo screening on a
system with many localized charges is not so surprising if we compare to quantum dots where the
Kondo effect is observed up to large numbers of electrons~\cite{Goldhaber-Gordon-1998}.
Nevertheless, the particular case of a 1D chain of localized charges in the Kondo regime still
requires theoretical investigations.

This analysis is consistent with the interpretation given in Ref.~\cite{Iqbal-2013a} for similar
observations using a QPC with six surface gates to tune the channel length. Our SGM experiment
brings additional information on this effect, since scanning the tip around the QPC, laterally or
vertically, changes the shape, extension, and symmetry of the channel potential. The circular and
almost isotropic rings in Fig.~2c show that the localized states survive to all these potential
deformations. The regularity of the successive rings also suggests that this localization occurs
rather independently of disorder, though possible crystal pinning effects should be investigated in
the future.

\begin{figure}[b]
\begin{center}
\includegraphics[width=16cm]{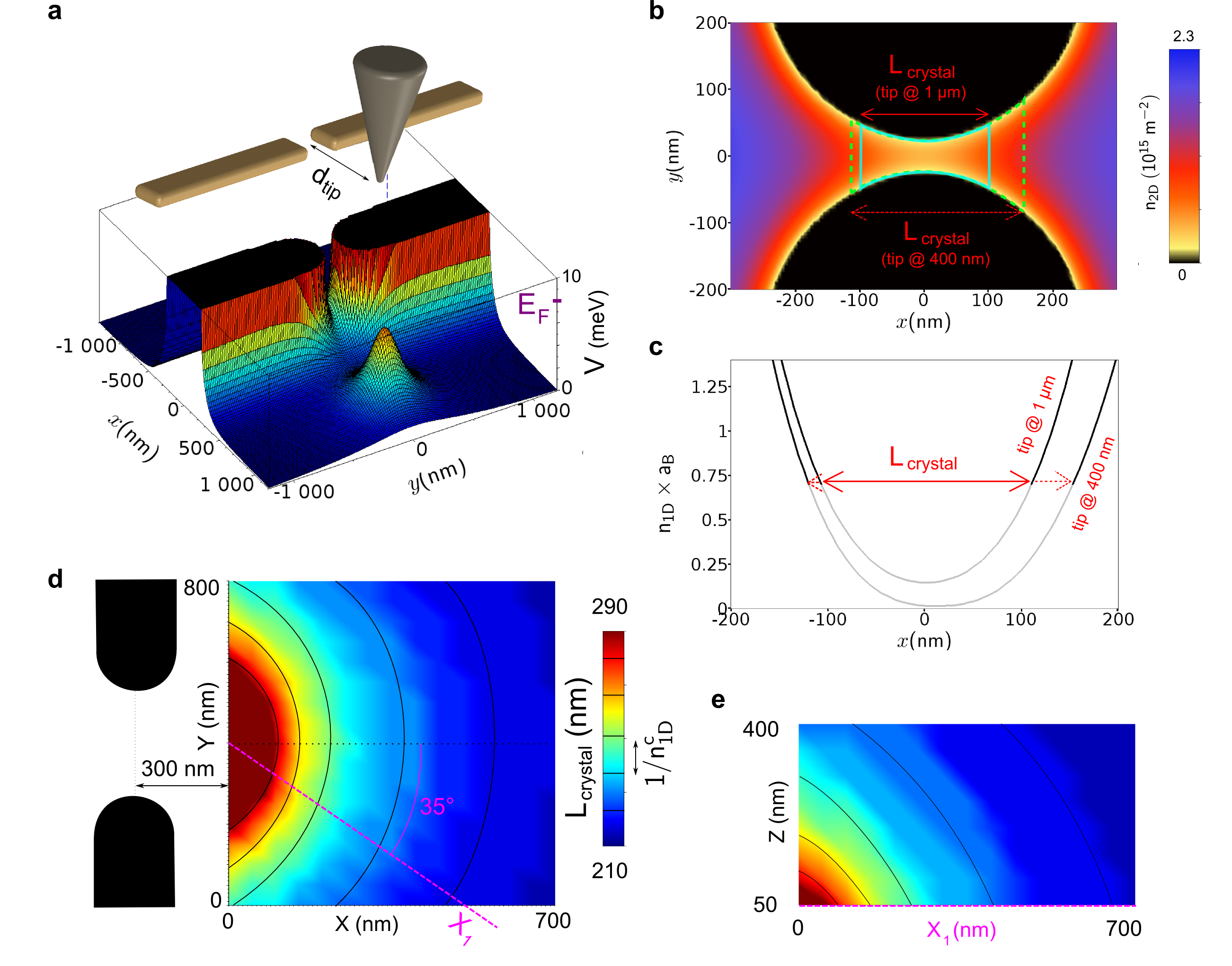}
\caption{\textbf{Calculation of the electron density and estimation of the Wigner crystal size.}
(a) Geometry of the metallic gates and SGM tip defined in the Comsol simulation software and
example of electrostatic potential map computed for a given gate voltage and tip position. The
Fermi energy $E_{\rm F}$ is 8~meV in the 2DEG. (b) Map of the two-dimensional electron density
$n_{\rm 2D}$ in the 2DEG computed classically but self-consistently with the potential, when the
tip is at 1~$\mu$m from the QPC. (c) One-dimensional electron density $n_{\rm 1D}$ obtained by
integration of $n_{\rm 2D}$ along the y-axis, when the tip is at 1~$\mu$m and 400~nm from the QPC.
Choosing a critical density $n_{\rm 1D}^{\rm c}=0.7/a_{\rm B}$ determines the expected size $L_{\rm
crystal}$ of the 1D Wigner crystal. (d) Computed size of the Wigner crystal as a function of tip
position in an horizontal plane 30~nm above the surface ($V_{\rm gate}=-1$~V). The region in red
corresponds to a closed contact (the electron density is zero at the QPC center for these tip
positions). Black lines indicate tip positions for which $L_{\rm crystal}$ is enlarged by $1/n_{\rm
1D}^{\rm c}$, corresponding at first order to the addition of one charge to the crystal. (e)
Computed size of the Wigner crystal for tip positions in a vertical plane (above line $X_1$ at
$35^{\circ}$ from QPC axis).}
\end{center}
\end{figure}

In Ref.~\cite{Iqbal-2013a}, the parameter controlling the number of localized states is the
effective length of the channel, defined in Ref.~\cite{Iqbal-2013} and computed using an analytical
approach assuming a fixed zero potential at the surface~\cite{Davies-1995}. This method is not
suitable to model our SGM experiment, as the tip is situated above the surface. To evaluate the
potential landscape in presence of the tip, we perform 3D classical electrostatic simulations in
the Thomas-Fermi approximation (see Methods and Supplementary Note 2) and compute self-consistently
the local potential $V(x,y)$ in the 2DEG and the local 2D electronic density $n_{\rm 2D}(x,y)$
(Fig.~5a). In this way, the tip-induced potential is correctly calculated, with the screening
effects from the 2DEG and the metallic gates taken into account. We obtain a good agreement between
calculated and experimental values regarding the gate voltage required to close the QPC, the tip
voltage to reach depletion in the 2DEG, and the cross-talk between the tip position and the QPC
opening. The effective channel length used in Ref.~\cite{Iqbal-2013a} was calculated in
Ref.~\cite{Iqbal-2013}, using the unscreened gate potential. This length cannot be calculated here
from our self-consistent potential, because screening effects induce non-parabolic transverse
confinement potentials.

We propose instead that the parameter controlling the number of localized charges is the size of
the region where the 1D Wigner crystallization should occur. This interaction-induced spontaneous
ordering is often discussed in terms of the Wigner-Seitz radius $r_{\rm s}=1/(2n_{\rm 1D}a_{\rm
B})$, representing the ratio of the Coulomb repulsion to the kinetic energy. A recent numerical
investigation of the 1D Wigner crystallization shows that the critical parameter $r_{\rm s}^{\rm
c}$ varies between 0.5 and 2, depending on the strength of the transverse confinement
potential~\cite{Shulenburger-2008}. To evaluate the size of the region where $r_{\rm s}$ is larger
than a given threshold, we calculate the 1D electron density by integration of the 2D electron
density in the transverse direction (Fig.~5b and 5c). As an example, we choose a critical value
$r_{\rm s}^{\rm c}=0.71$ corresponding to a critical density $n_{\rm 1D}^{\rm c}=0.7/a_{\rm B}$,
and evaluate the size $L_{\rm crystal}$ where the density is lower than $n_{\rm 1D}^{\rm c}$. This
size is found to vary from 210 to 290~nm when the tip is approached by 600~nm towards the QPC,
which shows that the tip can strongly affect the size of the low density region, and hence the
number of localized charges. The tip positions leading to the same $L_{\rm crystal}$ form rings
centred on the QPC, both for horizontal and vertical scanning planes (Fig.~5d and 5e), in the same
way as the conductance oscillations observed in the SGM experiment (Fig.~2c).

Our classical simulation holds only for an estimate of the size $L_{\rm crystal}$, but cannot be
used to calculate the number of localized charges, since quantum mechanics dominates at such a low
density. Note that charges in this crystal are not expected to be uniformly spaced, because the
potential of a QPC shows a strong curvature. This non-uniform situation would require an extension
of the concept of Wigner crystal which is usually studied in a flat potential landscape. A rather
crude approach to evaluate how many charges can be added by approaching the tip is to suppose that
one charge is added to the crystal each time the region is enlarged by $1/n_{\rm 1D}^{\rm c}$
(about 14~nm for $r_{\rm s}^{\rm c}=0.71$). With this assumption, about 5 charges can be added to
the crystal when the tip is approached close to the QPC (Fig.~5d). This value is qualitatively
consistent with the 4 oscillations observed in the experiment, and interpreted as the addition of 8
charges. Simulations also show that the number of charges can be modified simply by changing the
split-gate voltage (see Supplementary Figure 5). This could explain the ZBA splitting observed
above 0.7~$G_0$ in absence of the tip (Fig.~1c).

Our assumption that electrons form a 1D system in the low density region is justified \textit{a
posteriori} by the fact that only the first and second transverse modes are occupied over the
length $L_{\rm crystal}$. The presence of the second mode at the extremities of this region
indicates that the system is not strictly 1D, but theory still predicts the formation of a Wigner
crystal in the second subband of quasi-1D wires, forming a zigzag chain~\cite{Meyer-2007}, as
possibly observed in experiments~\cite{Hew-2009,Smith-2009}. Interestingly, the simulations show
that a small crystallized region survives when the second mode reaches the central part the
channel, which could explain the 0.7 analogues often observed between the first and second
conductance plateaus.

In summary, we observe a periodic modulation of the conductance anomalies in a QPC at very low
temperature while tuning continuously the potential with the polarized tip of a scanning gate
microscope. We explain this experimental observation by the formation of an interaction-induced
localized state in the QPC channel, which gives rise to a single- or two-impurity Kondo effect
depending on the odd or even number of localized charges, respectively. Indeed, electrostatic
simulations show that the electron density in the channel is low enough to result in a spontaneous
1D Wigner crystallization. Our study gives new information on QPC conductance anomalies, which
should guide future theoretical works, and will open the way to further experimental investigations
involving fine tuning of the QPC potential using various methods.

%------------------------------------------------------
\section{Methods}
%------------------------------------------------------

\subsection*{Sample and measurement}

The QPC is designed on a GaAs/AlGaAs heterostructure hosting a 2DEG 105~nm below the surface with
$2.5 \times 10^{11}$ cm$^{-2}$ electron density and $1.0 \times 10^6$ cm$^2$V$^{-1}$s$^{-1}$
electron mobility. A Ti/Au split-gate is defined by e-beam lithography on a mesa with four ohmic
contacts and forms a 270~nm long and 300~nm wide opening. The device is fixed to the mixing chamber
of a dilution fridge, in front of a cryogenic scanning probe
microscope~\cite{Hackens-2010,Martins-2013a,Martins-2013b}. The QPC is cooled down to a base
temperature of 20~mK at zero gate voltage. The four-probe differential conductance $G={\rm d}I/{\rm
d}V_{\rm bias}$ is measured by a standard lock-in technique, using a 10~$\mu$V AC excitation at a
frequency of 123~Hz. A series resistance of 600~$\Omega$ is subtracted from all data, in order to
have the conductance of the first plateau at $2e^2/h$. Since the temperature evolution of the
zero-bias peak does not saturate below 90~mK, the temperature of electrons in the QPC is probably
below this value.

\subsection*{Scanning gate microscopy}

The tip of a commercial platinum-coated cantilever is fixed on a quartz tuning fork, which is
mounted on the microscope actuators. The position of the QPC is determined by SGM, as the tip
position corresponding to the maximum change in conductance while scanning at large tip-surface
distance. Then, the tip is lowered to a few tens of nanometres above the surface and scanned at
fixed height on a single side of the 200~nm thick split-gate in the scanning area shown in
Fig.~1a. All the SGM results reported here are obtained for a tip voltage of $-6$~V and a
tip-to-surface height of 40~nm (except for vertical scans in Fig.~2c starting at 30~nm). Note that
the dilution fridge stays at its base temperature of 20~mK during tip scanning.

\subsection*{Electrostatic simulations}

Classical electrostatic simulations are performed with the Comsol software. We model the system in
three dimensions as follows. The 2DEG plane is located 105~nm below the surface according to our
heterostructure. The region between the 2DEG and the surface is filled with the GaAs dielectric
constant $\epsilon_r=12.9$. The initial electron density in the 2DEG is set at $2.5 \times 10^{11}$
e$^-$cm$^{-2}$ by the addition of a uniform plane of positive charges modelling ionized dopants (in
the same plane as the 2DEG for better computation stability). The metallic gates are 120~nm thick
and define a 270~nm wide and 300~nm long constriction, corresponding to our sample geometry. The
tip is modelled by a cone with a $30^{\circ}$ full angle and a 30~nm curvature radius at the apex.
The tip voltage is fixed at $-6$~V as in the experiment. For a given choice of gate voltage and tip
position, the local potential and density are computed self-consistently by successive iterations.
These calculations therefore include screening effects in the 2DEG.

%------------------------------------------------------
\section{Acknowledgements}
%------------------------------------------------------

We thank H. Baranger, J. Meyer, X. Waintal, D. Weinmann, J.-L. Pichard, and S. Florens for
discussions. This work was supported by the French Agence Nationale de la Recherche (``ITEM-exp''
project), by FRFC grant $n^o$ 2.4503.12, and by FRS-FNRS grants $n^o$ 1.5.044.07.F and J.0067.13.
F.M. and B.H. acknowledge support from the Belgian FRS-FNRS, S.F. received support from the FSR at
UCL, and V.B. acknowledges the award of a ``chair d'excellence'' by the Nanosciences foundation in
Grenoble.

%------------------------------------------------------
\section{Author contributions}
%------------------------------------------------------

B.B. and F.M. performed the low-temperature SGM experiment with the assistance of S.F., B.H., and
V.B.; B.B. and H.S. analyzed the experimental data and wrote the paper; A.C., A.O., and U.G. grew
the GaAs/AlGaAs heterostructure; C.U. and D.M. processed the sample; S.F., B.H., and F.M. built the
low temperature scanning gate microscope; G.B. performed the electrostatic simulations; B.B., F.M.,
S.F., B.H., U.G., D.M., S.H., G.B, V.B., M.S., and H.S. contributed to the conception of the
experiment; all authors discussed the results and commented on the manuscript.

%------------------------------------------------------

%------------------------------------------------------

%------------------------------------------------------
% Supplementary Information
%------------------------------------------------------

\newpage
\setcounter{figure}{0}
\renewcommand{\thefigure}{S\arabic{figure}}
\hrulefill \vspace{1cm}
\begin{center}
{\LARGE Supplementary Information}
\end{center}
\vspace{1cm} \hrulefill 

%------------------------------------------------------

\vspace{4cm}
\begin{figure}[h!]
\begin{center}
\includegraphics[width=15cm]{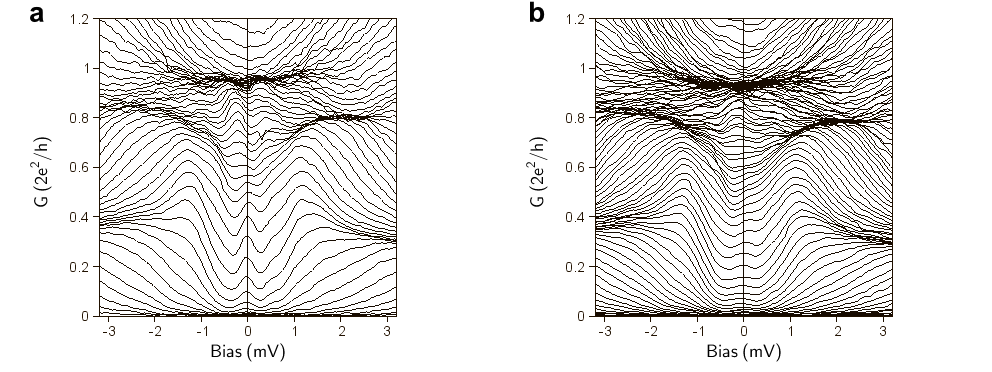}
\caption{\textbf{Supplementary Figure 1 : Zero-bias anomaly (ZBA) at different temperatures.} (a)
Differential conductance $G$ versus source-drain bias for different gate voltages at a base
temperature $T=20$~mK. (b) Same as in (a) at $T=900$~mK. Both data are recorded when the tip is far
from the surface. The zero-bias anomaly is strongly reduced in (b) as compared to (a).}
\end{center}
\end{figure}

%------------------------------------------------------

\newpage
\begin{figure}[h!]
\begin{center}
\includegraphics[width=17cm]{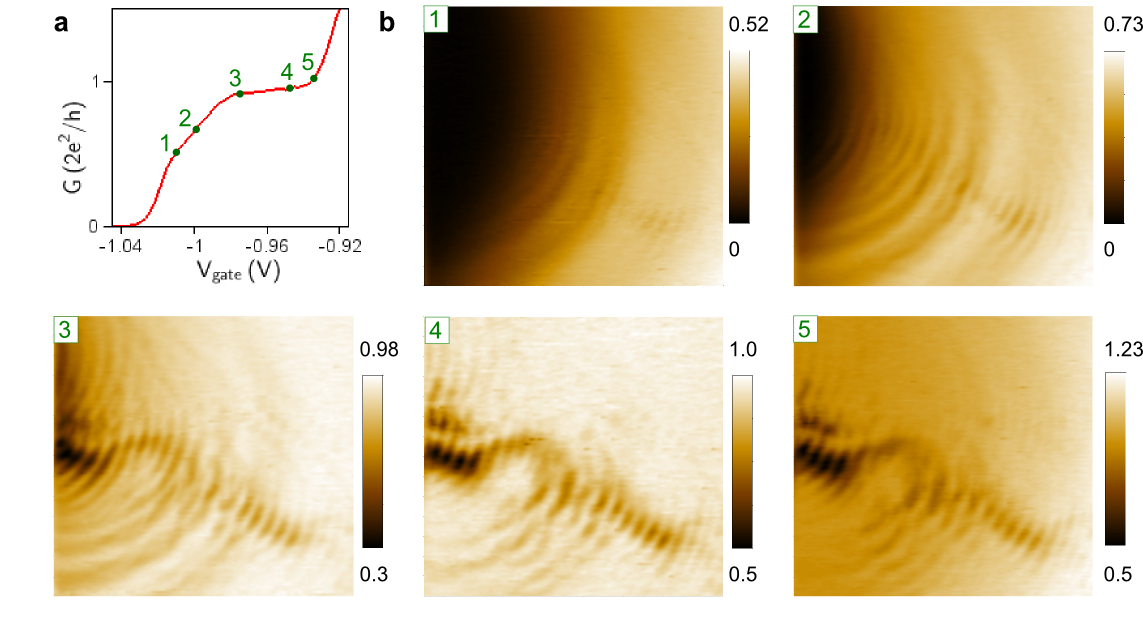}
\caption{\textbf{Supplementary Figure 2 : Scanning gate microscopy for different QPC openings.} (a)
Linear conductance $G(V_{\sf gate})$. (b) SGM maps of the conductance $G$ versus tip position
$(X,Y)$ in units of $2e^2/h$, recorded at different QPC openings: the gate voltage $V_{\sf gate}$
is $-1.010$~V, $-1.000$~V, $-0.975$~V, $-0.950$~V, and $-0.940$~V, for maps labelled 1 to 5,
respectively. It is clearly seen that the concentric rings exhibit a stronger contrast below the
first plateau, whereas only electron flow and interference fringes remain when the first mode is
fully open. These two types of conductance oscillations are superimposed on a global gating effect,
which appears as a global decrease in conductance as the tip approaches towards the QPC (in all
maps except in 4).}
\end{center}
\end{figure}

%------------------------------------------------------

\newpage
\begin{figure}[h!]
\begin{center}
\includegraphics[width=15cm]{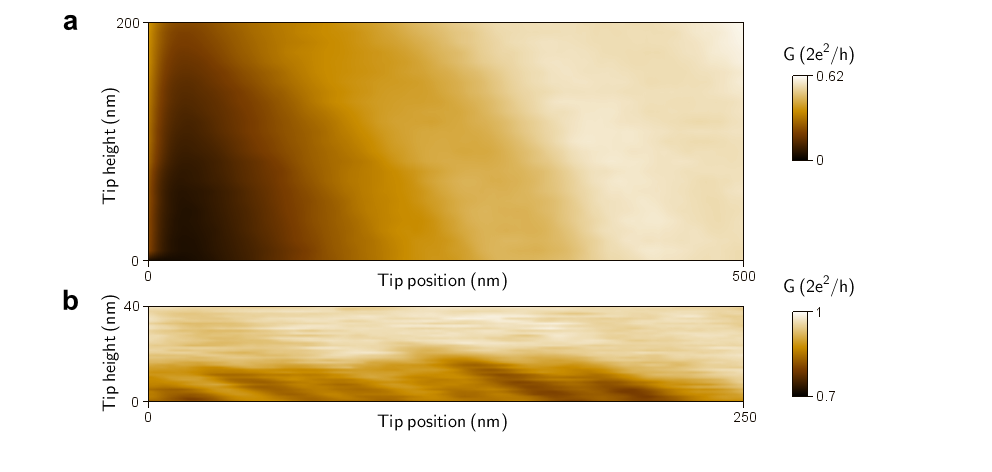}
\caption{\textbf{Supplementary Figure 3 : Scanning gate microscopy images in a vertical plane.}
(a,b) Conductance map $G(X,Z)$ recorded by scanning the tip in a vertical plane along lines 1 and 2
(defined in Fig.~2c of the paper), at gate voltage $V_{\sf gate}=-1$~V and $-0.95$~V, respectively
(vertical scans start from 30~nm above the surface). The origin of the ring structures appears
clearly from these vertical SGM maps. As the tip is scanned higher, the conductance oscillations in
(a) bend towards the QPC channel and are still visible when the tip is lifted by 200~nm: this
behaviour is consistent with a direct electrostatic effect. On the contrary, the interference
fringes in (b) disappear when the tip is lifted more than 20~nm (for this tip voltage of $-6$~V):
when the tip is too far from the 2DEG, the tip-induced potential no longer depletes the 2DEG, which
is a necessary condition for electrons to be backscattered, and hence for interferences to show up.}
\end{center}
\end{figure}

%------------------------------------------------------

\newpage
\begin{figure}[h!]
\begin{center}
\includegraphics[width=17cm]{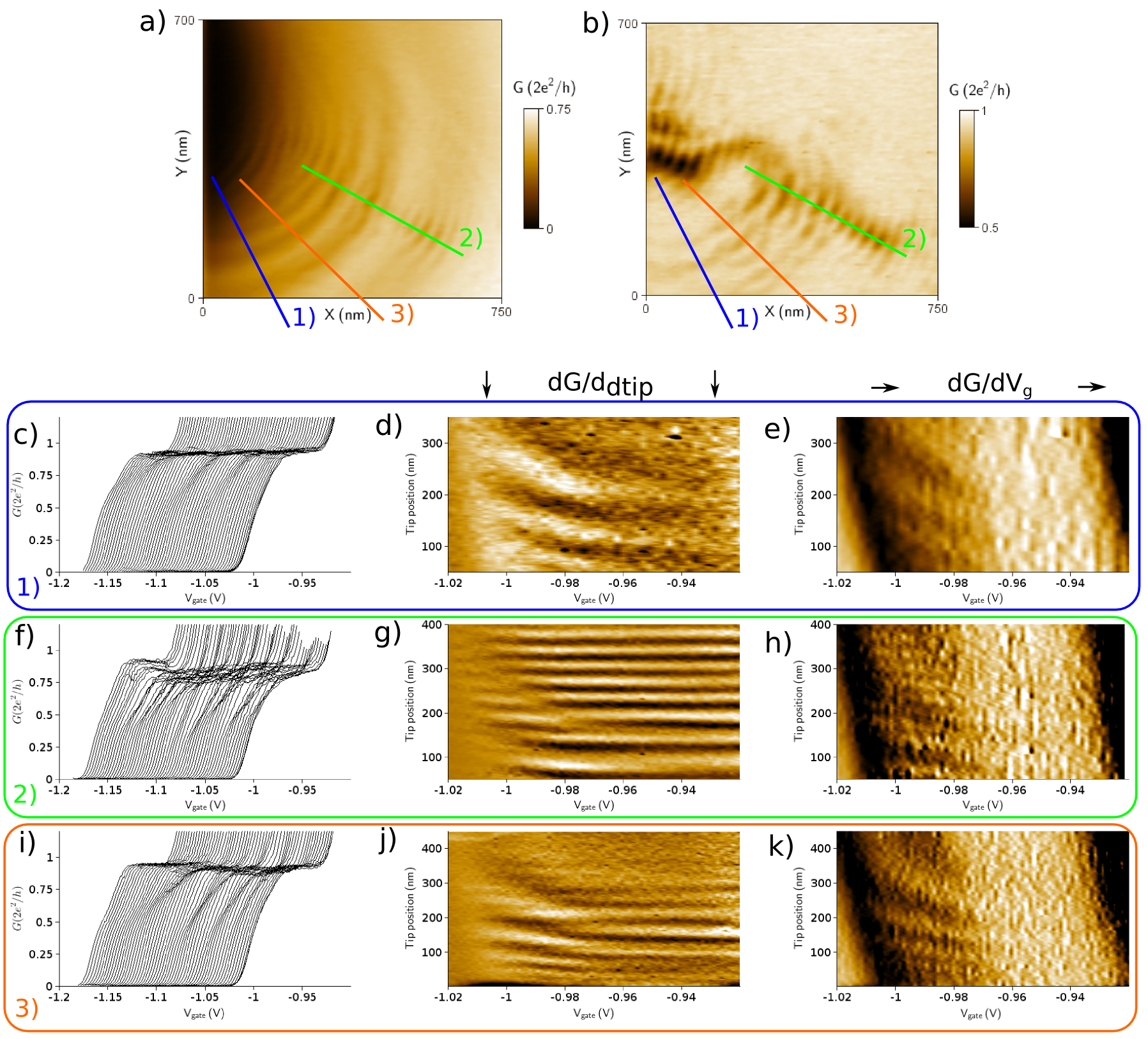}
\caption{\textbf{Supplementary Figure 4 : Concentric rings and interference fringes.} The figure
presents SGM data similar to those in Fig.~3 of the paper, but recorded along 3 different lines.
The lines are indicated in the SGM maps (a) and (b) for QPC openings below and on the first
plateau, respectively. For each line, the data are presented in 3 different ways. (c,f,i) Traces
$G(V_{\sf gate})$ for several tip positions along the chosen line (curves are shifted in $V_{\sf
gate}$ for clarity). (d,g,j) Color-plot of the same data, differentiated with respect to tip
position along these lines ($\partial G/\partial d_{\sf tip}$). (e,h,k) Color-plot of the same
data, differentiated with respect to gate voltage ($\partial G/\partial V_{\sf gate}$). The figure
is discussed in details in Supplementary Note 1.}
\end{center}
\end{figure}

%------------------------------------------------------

\newpage
\begin{figure}[h!]
\begin{center}
\includegraphics[width=15cm]{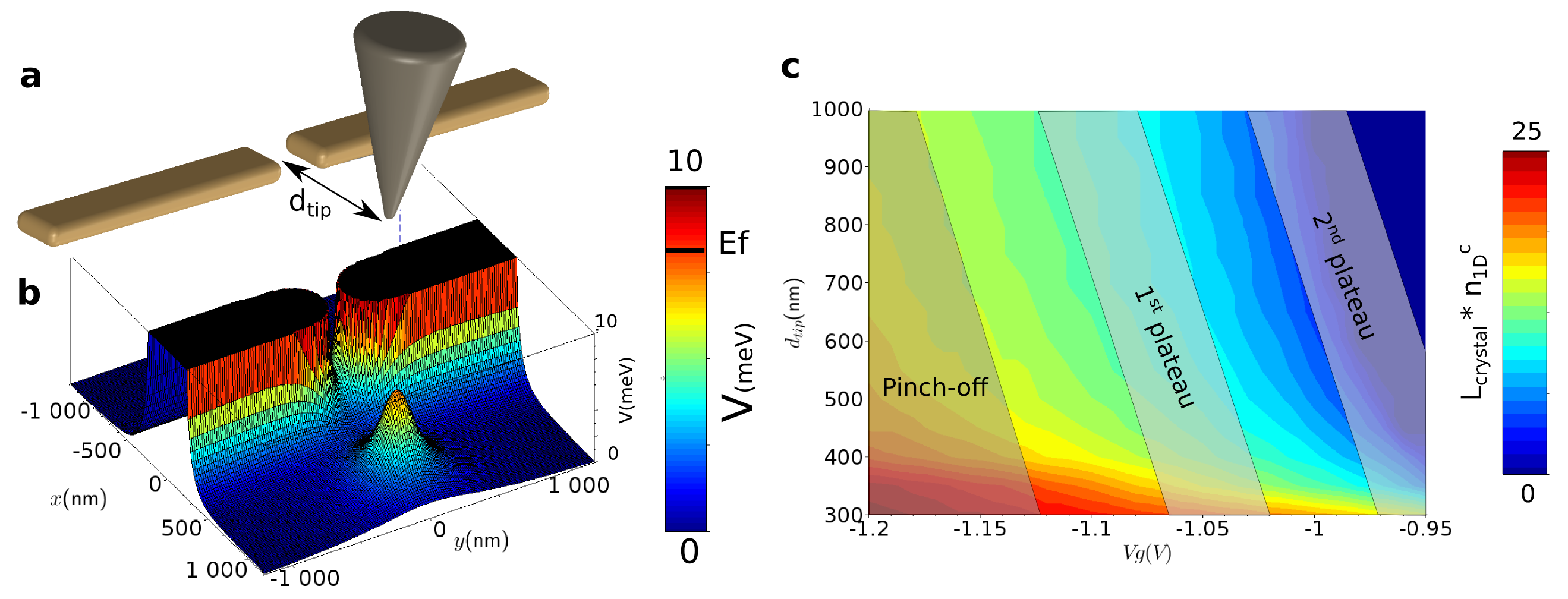}
\caption{\textbf{Supplementary Figure 5 : Electrostatic simulation of the QPC electron density.}
(a) Metallic split-gates and SGM tip defined in the Comsol software. (b) Self-consistent
electrostatic potential map computed classically for a given gate voltage and tip position. (c)
Estimated number of localized charges $L_{\sf crystal} \times n_{\sf 1D}^{\sf c}$ versus gate
voltage and tip distance (from the QPC centre). The figure is discussed in details in Supplementary
Note 2.}
\end{center}
\end{figure}

%------------------------------------------------------

\newpage
\noindent\textbf{Supplementary Note 1 : Conductance oscillations and interference fringes.}

In Supplementary Figure 4, we present three sets of data, showing the conductance versus
split-gates voltage and tip position along three different scanning lines, as indicated in
Supplementary Figure 4a and 4b. These data are analyzed in different ways: raw data $G(V_{\sf
gate})$ for a set of tip positions, and two color-plots obtained by differentiating the signal with
respect to (w.r.t.) either tip position, or gate voltage. As the interferences and the rings do not
behave in the same way with respect to these two parameters (tip position and gate voltage),
differentiating the signal w.r.t. one or the other of these parameters does not highlight the same
details. The chosen line also affects which phenomenon is observed, as interferences are present
only in specific regions (i.e. in the branches).

Along the blue line 1, the interferences are almost absent, hence the signal shows only the
concentric rings. Supplementary Figure 4c shows that these rings correspond to successive
modulations of the 0.7 anomaly. These modulations appear clearly in Supplementary Figure 4d and 4e
where the conductance is differentiated w.r.t. $d_{\sf tip}$ and $V_{\sf gate}$, respectively.

Along the green line 2, the interferences form the main signal. Supplementary Figure 4f shows that
these interferences strongly affect the conductance of the plateau, whose value is (in average)
below $2e^2/h$, because of backscattering by the tip located above a branch. Differentiating these
data w.r.t. tip position in Supplementary Figure 4g highlights these interferences. However, the
ring structure is still present along this line, and differentiating the signal w.r.t. gate voltage
in Supplementary Figure 4h recovers signatures of the rings.

Finally, data presented in the paper are shown in Supplementary Figure 4i, corresponding to the
orange line 3. As interferences are slightly visible along this line in Supplementary Figure 4b,
the signal is a mix between the two behaviors described above. Differentiating the signal w.r.t.
tip position in Supplementary Figure 4j gives a rather unclear picture, where interferences and
rings are mixed. However, as the interferences do not accumulate much phase with gate voltage,
differentiating the signal w.r.t. gate voltage in Supplementary Figure 4k hides these interferences
and highlights only the rings structure, which behaves in a very different way.

This analysis shows that differentiating the signal w.r.t. some parameter shall be used really
carefully. For this reason, we used this differentiation technique only once in the paper
(Fig.~3b), in a demonstrative way to show that the rings structure is contrasted below the first
plateau, but not to perform any quantitative analysis.

%------------------------------------------------------

\newpage
\noindent\textbf{Supplementary Note 2 : Electrostatic simulation of the QPC electron density.}

In order to justify that a finite number of charges can localize in the QPC channel, we perform
self-consistent classical electrostatic simulations using the Comsol software, as explained in the
paper (see Fig.~5 and Methods). We present here additional details and results of these
simulations.

We model the system in three dimensions as shown in Supplementary Figure 5a. The metallic gates are
120\,nm thick and define a 270\,nm wide and 300\,nm long constriction, corresponding to our sample
geometry. The tip is modelled by a cone with a rounded apex of curvature radius 30\,nm. The 2DEG
plane is located 105\,nm below the surface according to our heterostructure and we define a
dielectric constant $\epsilon_r=12.9$ between the 2DEG and the surface to model GaAs. The doping
layer is represented by a uniformly charged plane, with a $2.5 \times 10^{11}$\,cm$^{-2}$ density
of ionized dopants, located in the 2DEG plane for stability reasons, and insensitive to the local
potential. When no voltage is applied neither on gates nor on the tip, the electron density in the
2DEG is $2.5 \times 10^{11} e^-$/cm$^2$. For non-zero gate and/or tip voltages, the local electron
density and the local electrostatic potential are computed self-consistently by successive
iterations. The computation therefore includes screening from the 2DEG.

An example of simulation is presented in Supplementary Figure 5b, showing the self-consistent
electrostatic energy $V(x,y)$ in the 2DEG, in presence of the tip located at 800\,nm from the QPC.
The pinch-off is obtained for a gate voltage of -1.17\,V (in absence of the tip), which is close to
the experimental value of -1.06\,V. The difference may come from the fact that we assume no charges
on the surface and that the doping layer is placed in the 2DEG plane rather than 40\,nm above. The
simulations show that -6\,V applied on the tip placed 30\,nm above the surface is the threshold
value for the depletion of the 2DEG below the tip, which is very consistent with our experiment.
Moreover, the rise in the QPC saddle point when the tip is approached by 500\,nm is about
800\,$\mu$eV in the simulation, compared to the 850\,$\mu$eV evaluated from Fig.~3b of the paper.

From the self-consistent 2D electron density, we calculate the 1D electron density by integration
along the transverse direction. The size $L_{\sf crystal}$ of the Wigner crystal formed inside the
QPC in the low density region depends on the critical density $n_{\sf 1D}^{\sf c}$, whose value is
still debated. As an example, we choose the value $n_{\sf 1D}^{\sf c}=0.7/a_{\sf B}$ and calculate
the size $L_{\sf crystal}$ as explained in the paper, for different tip positions and gate
voltages. Assuming in a first approximation that the number of localized charges is proportional to
$L_{\sf crystal}$ times the density $n_{\sf 1D}^{\sf c}$, we plot this number of charges in
Supplementary Figure 5c, with respect to gate voltage and tip distance (measured from the QPC
centre). The figure also indicates roughly the regions corresponding to the transitions between
conductance plateaus. These regions have been tilted by the average cross-talk of the tip on the
QPC opening (partially screened by the metallic split-gates) according to the experiment (Fig.~3b).

Supplementary Figure 5c shows that approaching the tip towards the QPC changes the number of
localized charges by several units, which is qualitatively consistent with the number of observed
oscillations in our experiment. A change by one crystallized charge corresponds experimentally to a
switch between a zero-bias peak and finite-bias peaks, i.e. to half a period of the oscillations.

It also shows that, when the tip is far ($d_{\sf tip}=1000$\,nm), the estimated number of localized
charges changes by about one unit during the first opening (along the $V_{\sf gate}$ axis). This is
consistent with the only splitting of the ZBA visible experimentally at a conductance around $0.7
\times 2e^2/h$ in Supplementary Figure 1a.

These results are also consistent with the experimental data reported in Ref.~\cite{Iqbal-2013-sup}.
Since increasing the channel length enlarges the crystallization region, the successive ZBA
splittings observed in their experiment could also be explained considering Wigner crystallization.

Finally, Supplementary Figure 5c shows that a finite number of charges can localize up to the
second plateau, which could explain the ``0.7 analogues'' often observed in QPCs: a structure
similar to the 0.7 anomaly at a conductance value of about $1.7 \times 2e^2/h$. It can also explain
data such as those presented in Fig.~1d of Ref.~\cite{Cronenwett-2002-sup}, where a zero-bias peak is
visible between the first and second plateaus, and may even present splitting in this range.

%\noindent\textbf{Supplementary references}

%------------------------------------------------------

\end{document}